\documentstyle[epsfig]{aipproc}
\begin{document}
\title{Radio Pulses from Cosmic Ray Air Showers}
\author{Trevor C. Weekes}
\address{Whipple Observatory, Harvard-Smithsonian Center for
Astrophysics, \\ P.O. Box 97, Amado, AZ 85645-0097.\\
e-mail: tweekes@cfa.harvard.edu}
\maketitle

\begin{abstract}
The first experiment in which radio emission was detected from high
energy particles is described. An array of dipoles was operated by
a team of British and Irish physicists in 1964-5 at the Jodrell
Bank Radio Observatory in conjunction with a simple air shower
trigger. The array operated at 44 MHz with 2.75 MHz bandwidth. Out
of 4,500 triggers a clear bandwidth-limited radio pulse was seen in
11 events. This corresponded to a cosmic ray trigger threshold of
5x10$^{16}$ eV and was of intensity close to that predicted. The
early experiments which followed this discovery and their
interpretation is described.
\end{abstract}

\section{Background}

It is a great honor to be asked to give the opening talk at this,
the first workshop on the radio emission from high energy
particles. I should stress that my involvement with the field was
ephemeral and that although I was the first one to actually see
evidence of radio emission from high energy particles, I have made
no contribution to the field for 34 years. This invitation
motivated me to open my Ph.D. dissertation, probably for the first
time since its completion \cite{weekes66}, and to realize how lucky
I was to be involved, albeit in a junior capacity, with some
outstanding physicists in a classic experiment in radiation
physics. Since I am now struck by how different the methods used in
this simple experiment are from those used by the contemporary
student I will belabor the experimental method and stress the
historical background as best as I remember it.

Experimental physics in the fifties was still very much conducted
in the shadow of post-war politics, economics and experience. The
positive contribution of physicists to the war effort ensured that
physics research received some support even in those austere times.
Still, ingenuity and creative solutions were necessary to accomplish
anything meaningful. As a graduate student in the Experimental
Physics Department of University College, Dublin, I was fortunate
to join Neil Porter's Cosmic Ray group (in 1962) which was building
a very high energy gamma-ray astronomy experiment. In those days
graduate students really did build experiments, perhaps because
little time was
spent at computer terminals. Porter had recently returned to
Dublin after a stint at the Atomic Energy Research Establishment
(AERE), Harwell in England and continued to collaborate with John
Jelley at AERE on these cosmic ray studies. It was a measure of the
times that our gamma-ray telescope consisted of ex-World War II
searchlight mirrors mounted on an old British-naval gun mounting and
that major support for the experiment came via a grant from the
U.S. Air Force. These were exciting times in high energy
astrophysics (as the field came to be called) with the first X-ray
sources yet to be discovered and pulsars and quasars still over the
horizon. The post-Sputnik explosion in physics research was still
in its infancy and its effects had not yet really reached the
British Isles.

The discovery of cosmic ray particle cascades, the so-called
Extensive Air Showers (EAS) \cite{auger38} opened a new era of
cosmic ray studies with most interest centering on using them to
explore the highest energy cosmic rays. The sharply falling
spectrum meant that the events of greatest interest were detected
infrequently by conventional means (arrays of spaced particle
detectors). In the wake of World War II and the availability of
radio equipment and expertise, Blackett and Lovell
\cite{blackett41}
were moved to attempt to detect EAS using radar techniques to
detect the ionizing trail left in the wake of the shower. These
experiments were unsuccessful (the lifetime of the ions was
overestimated) and the technique was not pursued. The feasibility
of using simple optical radiation telescopes to detect EAS was
demonstrated by Galbraith and Jelley \cite{galbraith53} following
the suggestion by Blackett \cite{blackett5*} that the combined
Cherenkov emission from all secondary cosmic rays in the atmosphere
might constitute 0.01\% of the night-sky light background. Since
the Cherenkov radiation was concentrated in the forward direction
with a lateral spread similar to that of secondary particles, this
technique did not offer too much for the detection of very large
EAS, particularly since it was limited to clear dark nights giving 
duty cycles of $<$ 10\%; it did however lead to the development of
an effective technique for very high energy gamma-ray astronomy
\cite{weekes00}. Jelley \cite{jelley58a} proposed that Cherenkov
radio emission at radio wavelengths might be detectable but at the
microwave
frequencies necessary to prevent destructive interference; the
predicted signal was small although potentially detectable
\cite{jelley58b}. The possibility that EAS might be detectable by
the fluorescent emission that they caused in the atmosphere was
explored, without success, by Greisen \cite{greisen6*}; it was
later
to be revived with great success by the University of Utah group
\cite{utah7*}.

In 1962 G. Askaryan published a short paper in Russian in which he
suggested that the particle cascade resulting from the interaction
of a high energy particle in a dense medium would not be
electrically neutral since the resulting positrons could decay in
flight; also the cascade would accumulate delta-rays and Compton
scattered electrons \cite{askaryan62}. Cherenkov radio emission
could then occur at longer wavelengths where there would be
coherent emission from the net negative charge. Askaryan was
primarily concerned with the emission of radiation from the
particle cascade that would result from the interaction of a cosmic
particle in a dense medium like rock, e.g., on the moon. Since this
dielectric material is essentially transparent to radio waves, it
can provide a large target mass for such elusive
particles as neutrinos. Askaryan also pointed out that geomagnetic
effects might also contribute to separation of the charged
components in the shower and this dipole might provide an
additional emission mechanism. He estimated the negative excess,
$\epsilon$ in rock to be about 10\%. If N = the number 
of particles in the
shower, then instead of incoherent emission from N particles, we
must consider coherent radiation from ($\epsilon\cdot$N)$^2$. If
N=10$^6$ and $\epsilon$=0.1, then the coherence factor is 10$^4$,
a huge gain. The
coherence condition would require that the dimensions of the
shower-emitting region should be comparable with the wavelength.
The Cherenkov emission is proportional to $\nu.d\nu$, so that the
reduced Cherenkov emission at, say, 50 MHz compared with 5 GHz
would be more than compensated for by the increased coherence
factor.

The Askaryan paper appears to have gone unnoticed by
experimentalists until a follow-up paper by Alikanyan
\cite{alikanyan63}, on the emission of radio emission by high energy
particles, was noted by Neil Porter who was asked to write an
abstract for its publication in English. The paper referenced the
earlier Askaryan paper but it was not available in Ireland. Knowing
that Jelley had been interested in such phenomena, Porter sent a
copy of the Alikanyan paper to Jelley with a note to the effect
that it did
not seem a very feasible technique (Porter, private communication).
Jelley acquired the English version of the Askaryan paper and
realized that the negative excess offered new possibilities for the
radio detection of air showers. After an exchange of letters Jelley
and Porter agreed to attempt a simple experiment. Since neither of
them had access to radio telescopes they decided to enlist the help
of the extensive post-war British radio astronomy community. F.
Graham Smith, then at the Cambridge Radio Observatory, offered to
assist; as he was about to take up a position of Deputy Director at
the Jodrell Bank Radio Observatory, it was agreed that the
experiment should be done there. 

\section{The Experiment}

The experiment was to consist of a large area, broad
band, medium wavelength, wide angle radio telescope coupled to an
adjacent air shower array which would trigger the recording of the
predicted radio analog signal from high energy
showers at a reasonable rate. A radio telescope with these
parameters is best matched by a simple dipole array. It was
proposed to use the frequency reserved for the new BBC video signal
at 44 MHz where a relatively noise-free bandwidth of 2.75 MHz could
be achieved when the BBC transmitter was turned off. This was
approximately from 00.00 to 9.00 AM (no late night talk shows!) so
this was to be a night experiment with limited running time. I do
not recall whose idea it was to use this band but, in retrospect, 
it was key to the
successful outcome of the experiment. With the increased use of
the radio bands for communication it would not be possible to do
this today.

As the junior graduate student in the Dublin group I was assigned
the task of designing and building the EAS trigger that would be
used to signal the arrival of a large EAS. The design was based on
calculations using analytical models of shower particle
distributions. I believe these simple calculations (on our state-
of-the-art IBM 1620 computer) constituted the only involvement of
a computer in the experiment; computers were not involved in 
data taking or data analysis. 

The construction of the air shower array entailed a number of trips
to AERE where the array would be built and tested. AERE, as a large
British government research laboratory
with seemingly limitless resources, 
was a wonderland to a young Irish student.  It was
possible to put together a somewhat primitive but functional array
of Geiger counters in a matter of weeks. Geiger counters were
chosen because they were cheap and easily available. The simple
design called for three trays of Geiger counters at the corners of
an equilateral triangle of side 50~m; a trigger required a signal
from each tray within 0.5 microseconds. In each of the three trays
we
had counters of various sizes so we could get a rough determination
of the particle density and hence the shower size and its
proximity. Geiger counters are slow and noisy and precautions had
to be taken to ensure that the radio system was not susceptible to
pickup from the shower trigger. To achieve this we planned to run
the complete shower array from one battery supply, the radio
detectors from another. In June 1964 the array was operated for a
week at AERE (beside the Tandem Generator Building) where the
trigger rate was about 2.5 events per hour, roughly corresponding
to an EAS threshold of 10$^{16}$ eV. 

In July, 1964, the array was moved to the pastoral setting of
Jodrell Bank, some 20 miles from Manchester. There, under 
Graham Smith's direction, a dipole array was being assembled,
primarily by Bob Porter, a University of Manchester graduate student. 

With John Jelley and senior Harwell technician, John Fruin, I
assembled the small air shower array adjacent to the dipole array
and in the shadow of a 15~m radio telescope whose instrument shack
we commandeered. As with any new experiment in the field it was an
exciting time. Within a week we had the array up and running and
the Harwell pair left me to operate it for a couple of months. 

\begin{figure}
\centerline{\epsfig{file=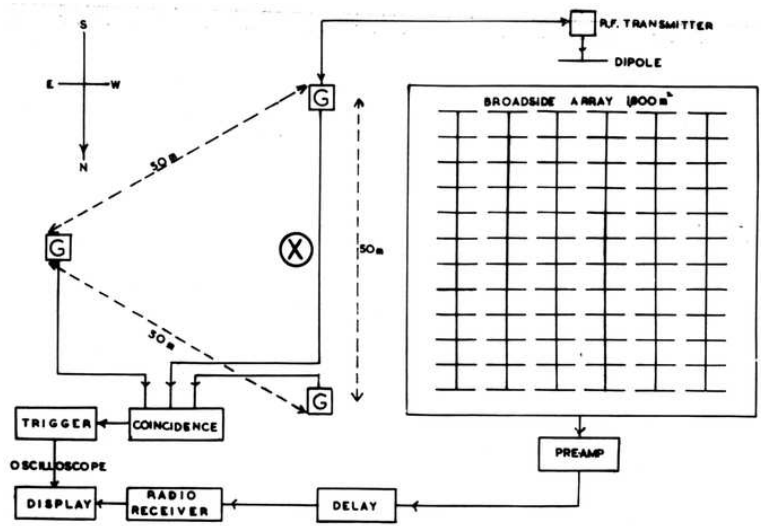,width=4in,angle=0.}}
\vspace*{0.5cm}
\caption{The Geiger counter array was set to one side of the dipole
array which had a total area of 1,700 m$^2$. The rf transmitter was
used in calibration experiments to measure the delay in the analog
signal \protect\cite{weekes66}.}
\label{array}
\end{figure}

The putative shower radio signal was delayed by passing it through
one kilometer of coaxial cable in a rack enclosed in a Faraday cage
with amplifiers distributed along the line to compensate for cable
attenuation. The output of the radio detector was to be displayed
on the 20 microsecond time-base of one channel of a two channel
oscilloscope; a time signal was displayed on the other. The scope
was triggered by the air shower array with the radio pulse expected
between 5.0 and 6.0 microseconds from the beginning of the trace
(there was considerable jitter in the array trigger)
(Figure~\ref{array}). This long
time-base gave a display of the radio background noise both before
and after the shower and was important in establishing that any
detected bandwidth-limited pulses were really associated with the
shower. The oscilloscope display was recorded on 35 mm film by a
Shackman camera; a small lens also brought into focus a clock and
hodoscope indicating which of the smaller Geiger counters had been
struck.

We quickly learned that radio pulse investigations operate in a
quite different environment from that seen by conventional radio
astronomy. The long integrations used by radio astronomers smoothed
out the man-made pulsed radio background which we soon discovered
to be dominated by such things as the nearby electric trains, the
typewriters in the administration building and the erratic ignition
systems of old cars driven by senior faculty. 
These problems were to be rediscovered by the
radio astronomers who within a few years would discover the radio
pulsars with very similar equipment.

Sadly I find I have no photographs of the dipole array which did
not lend itself to photography (no more than the air shower array
of Geiger counters did!). In contrast I have a detailed written log
of those days; I was scheduled to get married in September and out
of guilt at leaving all the marriage arrangements to my future
wife, I wrote to her every day; romantic soul that I was, I filled
my letters with day-to-day accounts of the progress of the
experiment!  
The Geiger array ran pretty much without problems (apart from the
inevitable eating of the high voltage cables by rabbits). The trays
and cables were 0.5 m above ground level and kept air tight; even
when the grassy area in which they were located was partially
flooded they continued to operate!   

The success of the experiment was in no small way due to the choice
of frequency and bandwidth and relative radio quietness at that
time. A chart recorder that recorded the integrated radio
brightness demonstrated the nightly passage of the Milky Way and
confirmed that we were indeed galactic noise-limited; the 
sky-brightness temperature varied from 6,000$^\circ$ to 20,
000$^\circ$. The receiver noise temperature was 450$^\circ$. No
form of automatic gain control was used. By phasing
the dipole array we were able to observe the transit of Cassiopeia
A and
thus determine our beamwidth (10$^\circ$ FWHM). This was my first
experience of real astronomy using non-visible wavelengths!

\section{The Results}

On August 19, 1964 all was ready for the first night of data
taking. John Jelley was a meticulous experimentalist and had left
a check list for the night operator to complete. Radio
observatories are often lonely places at night and it was a little
eerie setting up an experiment in a remote location to begin
operation after midnight. All was now ready to test the hypothesis
that air showers produced detectable radio emission. I confess to
some anticipation and excitement as I set the system to operate
that first night. These emotions were balanced by my discovery the
next morning that after the first event the camera had jammed and
no data was taken! The next night I was more careful: I double
checked the Jelley list and waited for an hour to be certain that
data was being recorded.

It was Sunday morning when I unloaded the camera, checked that the
various housekeeping chart recorders showed nothing unusual and
developed the long roll of film in the observatory dark room. In
the dark room lights I made a quick check on the film and was
delighted to see on the fifth image recorded exactly what we
had been looking for, evidence for a very large radio pulse at
precisely the right point on the time-base. Later that day I
examined the 
full roll of film and noted no other evidence for emission but no
background events either.

This was my first experience of a scientific discovery and young
that I was, I assumed this was the norm of scientific research and
that I should expect such happenings on a regular basis! Little did
I realize that (a) I would wait another twenty years for a
comparable happening (b) that this was the largest pulse we would
record and (c) there would be no other evidence for emission in the
remainder of the eight days of the run. Wisely I decided I should
celebrate immediately; I did not have access to a telephone on
site, there was
no e-mail or other way to communicate with my supervisors, so I
mounted my trusty government-issue bicycle and repaired to the
nearest village some miles away where I downed some good English
bitter. 

The next week was a disappointment. All systems seemed to run
perfectly but  no radio pulses were evident. We were running with
just one quarter of the dipole array completed. At the end of the
week we were faced with the big decision; did our preliminary run
justify the effort and expense of completing the array? I opted for
completing the array which was an easy decision on my part since I
was leaving for a few months and the cost, which was minimal, would
be borne by Jodrell Bank.

\begin{figure}
\centerline{\epsfig{file=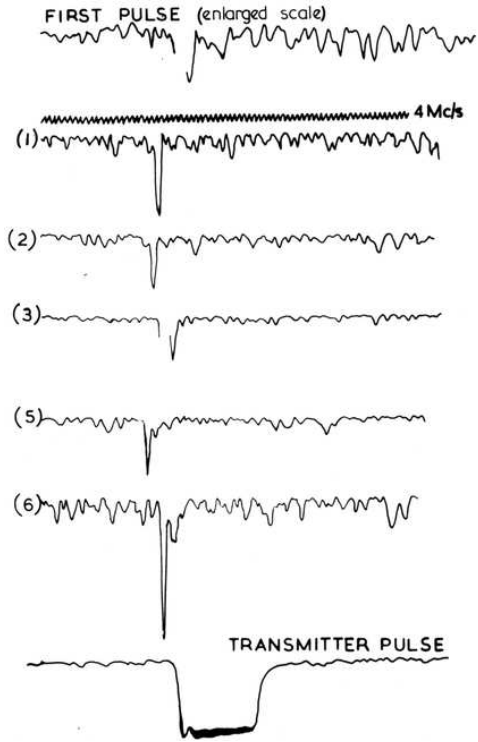,width=2.5in,angle=0.}}
\vspace*{0.5cm}
\caption{Some of the oscilloscope traces showing the "large" radio
pulses. The first (offscale) pulse was recorded on a different
timebase than the others. The arrival time of the calibration
transmitter pulse must be corrected for the time delay in the
Geiger circuitry \protect\cite{weekes66}.}

\label{pulses}
\end{figure}

In a month the array was completed and the experiment was run on 
almost every night until March 1965 (I did another stint of
operation in December, 1964). To my immense relief more pulses
appeared but only one off-scale pulse. In some 4,500 air
shower triggered events, there were eleven with clearly discernible
pulses in the right delay window. There were no comparable pulses
anywhere else on the event times-bases or on time-bases
triggered by a clock every half-hour. Some of these events are
reproduced in Figure~\ref{pulses}.

\begin{figure}
\centerline{\epsfig{file=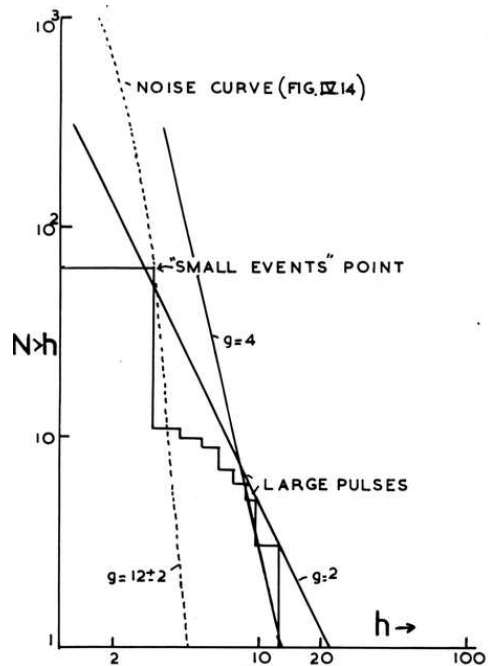,width=2.5in,angle=0.}}
\caption{The distribution of pulse sizes measured in the delay
window where radio air shower pulses are expected is plotted with
a single point for the observed small events. The normal receiver noise
spectrum pulse height distribution is also shown \protect\cite{weekes66}.}
\label{spectrum}
\end{figure}

We also performed an analysis on the traces
 that did not show
an obvious large pulse. This analysis was done in Dublin and
consisted of recording the position of the maximum ''noise"
fluctuation on each oscilloscope trace. A clear peak was seen in
the distribution in the anticipated interval (and none in the clock
triggered events) providing independent evidence for the detection
of radio emission. This data was folded with the distribution of
measured pulse sizes in the large events to give the size spectrum
shown in Figure~\ref{spectrum}.

\section{Interpretation}

The observed rate of detected radio pulses
was consistent with emission from cosmic ray air showers initiated
by primaries of energy 5x10$^{16}$ eV. The energy in the received
radio pulse was about 1 eV. The air shower array did not give any
information about the shower arrival direction and the small
counter triggers did not allow a precise definition of either the
size of the shower or the position of its axis. A series of 
follow-up experiments involved (a) use of an optical air shower to
trigger
on smaller showers with known arrival directions (b) operation in
coincidence with the 15~m radio telescope fitted with a radio
receiver at 150~MHz (c) a short run with one section of the dipole
array rotated in orientation to favor North-South polarization.
After 1966 (when I left UCD) the Dublin group extended the observed
radio pulse frequencies into the UHF band and  
detected  radio pulses that were probably radiated incoherently
 \cite{fegan70}.
None of these experiments gave conclusive results on the nature of
the emission but all were consistent with the hypothesis that the
radio emission originated in large air showers.

Alternative radiation hypotheses were also considered including:
(a) nuclear field bremsstrahlung; (b) transition radiation; (c)
induction effects; (d) cosmic rays striking receivers; (e)
reflection of TV signal; (f) molecular emission. Preliminary
estimates eliminated all these as possible explanations of the
observed radio emission \cite{jelley65a}\cite{weekes66}.
 
In the original Askaryan paper the possibility that radio emission
might also result from geomagnetic effects on the shower particles
were mentioned. This concept was developed by Porter in Dublin as
we built the experiment. Two mechanisms were recognized in addition
to the Cherenkov radiation from the negative excess: i) the
separation of the positive and negative charges as they traversed
the earth's magnetic field would create a dipole and give rise to
radio emission; ii) as shower develops in the atmosphere, electrons
are being moved to one side, positrons to the other by the magnetic
field; the result is that the shower constitutes a current element
which will also radiate radio waves. The  full treatment of these
processes was done eventually by Kahn and Lerche at Manchester
University \cite{kahn66} and Colgate \cite{colgate67} at Los Alamos.
A good summary of all aspects of these early experiments
can be found in a review article by Allen \cite{allen71}. Remarkably
all three processes seemed to give signals of comparable magnitude
(but with the detection of polarization and the measured
distribution of radio signal with frequency, the consensus now
seems to be that the current element is the dominant emission
mechanism).  

\section{Conclusions}

The results from the initial experiment were published in a Letter
to Nature \cite{jelley65a} and in a Nuovo Cimento paper
\cite{jelley66}. They were also reported at the 9th International
Conference on Cosmic Rays (ICRC) in
London in 1965 \cite{jelley65b}. By that time the
phenomenon had been independently confirmed by an experiment
performed by the UCD group in the Dublin mountains, using a helical
antenna with a receiver at 70 MHz with bandwidth 20 MHz and a
plastic Cherenkov detector shower trigger \cite{porter65}. These
early results were given some prominence in the report of the
ICRC EAS Highlight speaker, K. Greisen \cite{greisen65}, from whose
paper the
following is quoted: ''The technique is barely in its infancy, and
too little is known about it to justify elaborate predictions.
However, we feel confident that this achievement is a significant
breakthrough, and that further study will reveal ways of obtaining
types of information about the showers that were not available by
other means. The signal will not, of course, be sensitive to fine
details like the structure of the shower core; nor does the method
seem adaptable to surveying at the same time all directions in
which showers may arrive. However, it appears that the method may
offer new orders of angular resolution; and it may complement
particle detectors by being sensitive to the conditions of showers
far above ground level. Also it will be able to detect showers in
steeply inclined directions, in which the particles are absorbed
before reaching the ground. Time will probably reveal other
possibilities that are not apparent at this early stage." Sir
Bernard Lovell was to note that this experiment fulfilled the
original purpose of the first telescope constructed at Jodrell Bank
in 1941 \cite{lovell8*}.

\begin{figure}
\centerline{\epsfig{file=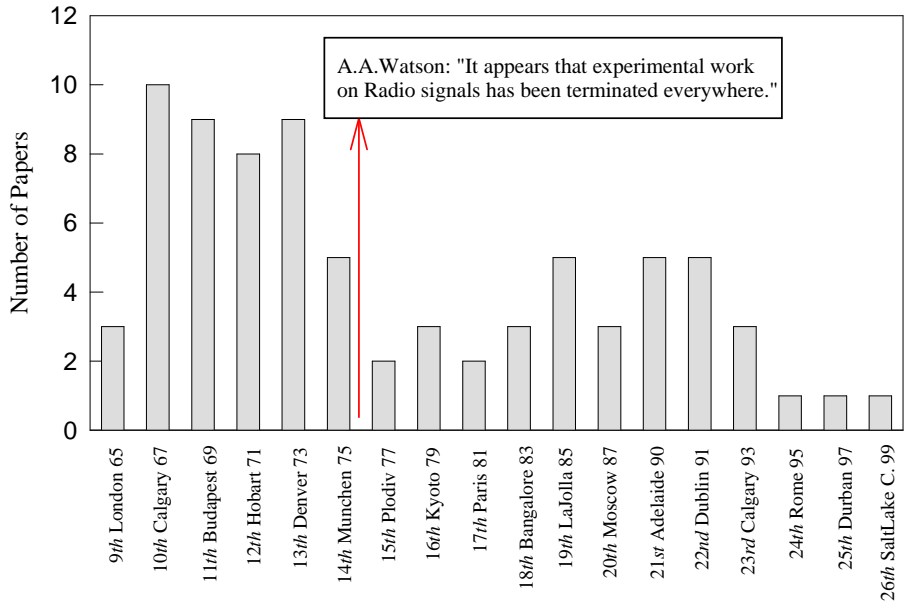,width=5.5in,angle=0.}}
\caption{The number of papers on the radio emission of air showers
presented at International Cosmic Ray Conferences from 1965 to the
present day (compilation by H. Badran).}
\label{badran}
\end{figure}

This optimistic assessment led to a spate of experimental activity
which was mostly aimed at an exploration of the radiation
mechanisms. Interest in the field can be assessed in the numbers of
papers presented on the technique in subsequent ICRCs 
(Figure~\ref{badran}).
Early interest in the potential of the technique seems to have
waned as it was recognized that the emission was directed and thus
not suited to distant shower detection and that the man-made pulse
radio
background was noisy and becoming more so. Thus the early hope that
radio detectors might be used in stand alone systems to detect
distant very large air showers was not realized. The recent
interest in the detection of showers of energy $>$ 10$^{20}$ eV has
renewed interest in the radio detection technique.

I am grateful to Peter Gorham and David Salzberg for allowing me to
indulge in this reminiscence of the early work and to Neil Porter
and David Fegan for jogging my memory on important details.

\end{document}